\documentclass[pre,aps,reprint,graphicx,amsmath,amsfonts,showpacs,showkeys]{revtex4-1}

\usepackage{algpseudocode}
\usepackage{bm}
\usepackage{color}
\usepackage{fixmath}
\usepackage{graphicx}

\draft 

\DeclareMathOperator{\Ord}{O}
\DeclareMathOperator{\mmean}{mean}
\DeclareMathOperator{\var}{var}

\DeclareMathOperator{\Pois}{Pois}
\DeclareMathOperator{\VST}{VST}
\DeclareMathOperator{\VSTL}{VSTL}

\newcommand{\ifterm}{\mbox{\textit{terminate}}}

\newcommand{\Statexi}{\Statex\hspace{\algorithmicindent}}

\begin{document}


\title{The symmetries of image formation by scattering. II. Applications} 


\author{Peter Schwander}
\thanks{These authors contributed equally.}
\author{Chun Hong Yoon}
\author{Abbas Ourmazd}
\email[To whom correspondence should be addressed. Email: ]{ourmazd@uwm.edu}
\affiliation{Dept.~of Physics, University of Wisconsin Milwaukee, 1900 E. Kenwood Blvd, Milwaukee, WI 53211}

\author{Dimitrios Giannakis}
\thanks{These authors contributed equally.}
\affiliation{Courant Institute of Mathematical Sciences, New York University, 251 Mercer St, New York, NY 10012}


\date{\today}

\begin{abstract}
We show that the symmetries of image formation by scattering enable graph-theoretic manifold-embedding techniques to extract structural and timing information from simulated and experimental snapshots at extremely low signal. The approach constitutes a physically-based, computationally efficient, and noise-robust route to analyzing the large and varied datasets generated by existing and emerging methods for studying structure and dynamics by scattering.  We demonstrate three-dimensional structure recovery from X-ray diffraction and cryo-electron microscope image snapshots of unknown orientation, the latter at 12 times lower dose than currently in use.  We also show that ultra-low-signal, random sightings of dynamically evolving systems can be sequenced into high quality movies to reveal their evolution.  Our approach offers a route to recovering timing information in time-resolved experiments, and extracting 3D movies from two-dimensional random sightings of dynamic systems. 
\end{abstract}


\maketitle 

\section{Introduction}
\label{secIntroduction}
In an earlier paper~\cite{GiannakisEtAl11a}, hereafter referred to as Paper~I, we presented a theoretical framework for 
analyzing snapshots formed by scattering.  In this paper, we demonstrate the power of this approach to 
reconstruct three-dimensional (3D) models and time-series from random sightings at extremely low signal, with no orientational or 
timing information.  The theoretical framework in Paper~I represents the information content of an ensemble of 
snapshots as a Riemannian manifold, and shows that the properties of operations in space give rise to 
object-independent symmetries.  Purposeful navigation on this manifold is tantamount to reconstructing a 3D
model of the sighted system and/or its evolution, in the sense that given any snapshot, any other can be 
produced on demand.  The symmetries of the manifold reveal its natural eigenfunctions, thus allowing 
physically-based interpretation of graph-theoretic analysis, and enhanced noise 
discrimination.  Simple algorithms then suffice to reach exceptionally low signal-to-noise levels unmatched 
by other approaches in terms of computational cost, noise robustness, or both.  As examples, we demonstrate structure recovery from radiation-sensitive objects at doses at least an order of magnitude below current levels (signal-to-noise ratio (SNR): $ -16\text{ dB} $), and reconstruction of time-series at SNR values as 
low as $ -21\text{ dB} $.  The versatility of the approach is demonstrated in the context of simulated and experimental data from X-ray diffraction, cryo-electron microscopy (cryo-EM), and optical snapshots using a variety of 
graph-theoretic techniques.  These applications demonstrate the generality of the symmetry-based
approach, elucidating at the same time the measures needed to deal successfully with experimental data, a key benchmark of the practical utility of any theoretical framework.

This paper is organized as follows.  Without claim to be comprehensive, Sec.~\ref{secPreviousWork} briefly summarizes previous work in the field to provide a context for the applications discussed in this paper. For the convenience of non-mathematical readers, Sec.~\ref{secConceptualSummary} provides a conceptual outline of the theoretical framework developed in Paper~I. Sec.~\ref{secSimulatedData} describes 3D reconstruction from simulated diffraction snapshots of single biomolecules at the signal level expected from single molecules in upcoming experiments utilizing the new generation of X-ray Free Electron Lasers (XFELs)~\cite{ShneersonEtAl08,FungEtAl08,SchwanderEtAl10a}.  Sec.~\ref{appSOD} establishes, in principle, the applicability of our approach to crystalline samples. Sec.~\ref{appCryo} addresses structure recovery from simulated and experimental cryo-EM snapshots of single molecules. In this case, essential experimental issues such as defocus variation must be faced and incorporated into the theoretical formalism.  Sec.~\ref{appPirouettePasDeDeux} demonstrates reconstruction of time-series (movies) from random sequences of ultralow-signal optical snapshots.  The paper concludes in Sec.~\ref{secConclusions} with a summary of our key findings and their implications.  Detailed points of a technical nature are elucidated in appendices, and movies provided as supplementary online material EPAPS~\cite{EPAPS}. 

\section{Previous work}
\label{secPreviousWork}
As described in Paper~I, we are concerned with constructing a model from random sightings of a system viewed in some projection,  i.e., by accessing  a  limited  number  of variables  describing  the state  of the  system.  A 3D model of an object and its evolution, for example, can be constructed from an ensemble of low-signal 2D snapshots without orientational information~\cite{Frank02,ScheresEtAl07,FungEtAl08,LohElser09,SchwanderEtAl10a,FischerEtAl10}. Modern graph-theoretic algorithms can now be used to discover low-dimensional manifolds representing the information content of datasets in some high-dimensional space determined by the measurement apparatus~\cite{TenenbaumEtAl00,RoweisSaul00,BelkinNiyogi03,DonohoEtAl03,CoifmanEtAl05,CoifmanLafon06,CoifmanEtAl10,BishopEtAl98}.  The power of these methods stems from their generality, in the sense  that no assumptions are made as to the nature of the data or their internal correlations. 
This brings with it four major challenges: (1) Interpretation of the analysis results (``what physical variables do the manifold
dimensions represent?''); (2) Computational cost and scaling behavior on moving from simulated (``toy'') 
datasets to experimental measurements; (3) Robustness against noise, particularly of non-additive, non-Gaussian types; and (4) Incorporation of inevitable and/or desirable experimental factors (``utility of the theoretical framework in practice''.)   

These issues can be brought to focus in the context of the much-discussed problem of recovering the 
orientation of cryo-EM snapshots of faint biological objects. Direct graph-theoretic attempts to determine 
the orientation of snapshots from a synthetic object were abandoned at a signal-to-noise ratio (SNR) of $ \sim 2 $ dB, even though only additive Gaussian noise was included~\cite{CoifmanEtAl08}. Noting that graph-theoretic analyses 
often ``fail to solve the cryo-EM problem, because the reduced coordinate system that each of them obtains 
does not agree with the projection directions''~\cite{CoifmanEtAl10}, properties specific to cryo-EM images were used to 
extract information from the snapshots. Graphs were then constructed using this information in order to 
assign physical meaning to the outcome of the analysis.  Orienting low-signal cryo-EM snapshots by 
utilizing so-called (straight) common-lines identified primarily in simulated data with additive Gaussian 
noise has reached remarkably low SNR values~\cite{SingerEtAl10}. However, such assumptions, while justified under 
some circumstances, are not generally valid. Common-lines, for example, are present only when elastic 
single-scattering dominates, are straight only when the wavelength of the incident radiation is so short that 
the Ewald sphere can be replaced by a plane, and are compromised by defocus variations essential for 
reliable structure recovery by cryo-EM~\cite{Frank06}.   

Symmetry-based assignment of physical meaning to the outcome of graph-theoretic analysis of scattering data and its favorable computational consequences were addressed in Paper~I.  Here, we are concerned with ability of this theoretical framework to deal with noise and other important factors encountered in experimental datasets.  This determines the practical utility of an approach as much as theoretical elegance and computational efficiency.  Below, we demonstrate the utility of our symmetry-based approach by applying a number of manifold-embedding techniques to a variety of simulated and experimental datasets (see Table~\ref{tabApplication}). 

\section{Conceptual summary of theoretical framework}
\label{secConceptualSummary}
A snapshot formed on a 2D detector by scattered radiation from an object can be represented by a vector,
with the intensity values recorded at the $ n $ detector pixels as components (Paper~I, Fig.~1).   Object motion 
and/or evolution (dynamics) change the pixel intensities, causing ``the vector tips'' representing the 
ensemble of snapshots to trace out a surface --- a manifold --- in the $ n $-dimensional data space. The number of 
degrees of freedom available to the object determines the dimensionality of the manifold traced out.
Rotations of a rigid object in 3D, for example, result in a 3D manifold.   

The data manifold represents the totality of information about the object gathered by the detector in the 
course of an observation.  Learning is tantamount to understanding the properties of this manifold 
sufficiently to ``navigate'' on it.  Learning the manifold generated by object rotations, for example, is 
equivalent to constructing a 3D model of the object, because, starting from any 2D projection (point on the 
manifold) any other 2D projection can be found by navigation, with the shortest route corresponding to a 
geodesic. 

As we are initially concerned with constructing 3D models from 2D snapshots, we consider a formulation 
of scattering by a single object in Fourier space so as to concentrate on the effect of rotations. This circumvents issues such as rigid 
shifts, which would otherwise have to be corrected or incorporated as additional manifold dimensions.  
Rotation operations do not commute.  One must therefore consider the order in which they are 
performed.  This leads to a distinction between so-called left and right ``translations,'' where a rotation operator $ \mathsf{ T } $ is placed to the left or right of another rotation operator $ \mathsf{ R } $, i.e., $ \mathsf{ T } \mathsf{ R } $ vs.\ $ \mathsf{ R } \mathsf{ T } $.  A left translation can be thought of as an active rotation in 3D space of the incident beam-detector arrangement 
(frame rotation) after $ \mathsf{ R } $.  Similarly, a right translation corresponds to an active rotation of the object 
(object rotation) before $ \mathsf{ R } $. As $ \mathsf{ T } \mathsf{ R } \neq \mathsf{ R } \mathsf{ T } $, left and right translations must be considered separately. Each forms an SO(3) group, and the total set of possible operations to be considered corresponds to $ \text{SO(3)} \times \text{SO(3)} $. 

A key question is this: Which, if any, of these operations leave the distances on the manifold unchanged, 
i.e., which operations are ``invisible'' to an ant crawling on the manifold?  These operations would represent 
symmetries --- more precisely isometries --- of the manifold.  For a detector with circular symmetry, the distances on the 
manifold are invariant under beam-detector rotations about the beam axis.  This is obvious; a frame rotation 
about the beam axis rotates all the snapshots by the same amount about that axis without changing them.  
This leaves the distances on the manifold unchanged.  The process of image formation on a circularly-symmetric detector at right angles to the illuminating beam thus has $ \text{SO(2)} $ isometry, i.e., of all possible 
$ \text{SO(3)} $  frame operations, the $ \text{SO(2)} $  subset of rotations about the beam direction leave the distances on the manifold unchanged.  
This is related to the projection of a 3D object on the 2D detector, which is equivalent to a ``central slice'' 
through the diffraction volume in reciprocal space. 

Consider next the $ \text{SO(3)} $  set of operations corresponding to object rotations.  It turns out that the metric measuring distance on the manifold can be decomposed into a homogeneous part, which varies uniformly with object rotation, plus a residual term, which acts as a fingerprint of the object (see Paper~I Sec.~III~C). Considering the homogeneous part only, the total set of symmetries, is then $ \text{SO(2)} \times \text{SO(3)} $. The same set of symmetries appears in certain models of the universe in general relativity~\cite{Taub51,Hu73}, and is associated with well-known eigenfunctions familiar in the context of spinning tops in classical and quantum mechanics~\cite{BiedenharnLouck81}.

The key point here is that the knowledge of the manifold symmetries, which stem from the nature of operations in space, allows one to determine the leading-order properties of the manifold under a very general set of scattering scenarios, including its natural eigenfunctions.  Projection of noisy datasets on these eigenfunctions is tantamount to noise discrimination.  The components of a data point representing a snapshot can then be directly related to its orientation (see Paper~I Sec.~III~D). 
     
\section{Applications}
\label{secResults}
It has long been known that the use of problem-specific constraints can substantially increase computational efficiency \cite{LeCunEtAl90}. By combining wide applicability with class specificity, symmetries represent a particularly powerful example of such constraints. In Paper~I, we used the object-independent symmetries of image formation to  recover 3D structure from a large ensemble of simulated, noise-free diffraction snapshots with a computational complexity  $10^4\times $ higher than the state of the art.
Here we demonstrate the noise robustness stemming from exploiting the symmetries of image formation. Examples include orientation recovery, 3D reconstruction, and movie extraction from ultra-low-signal diffraction or image snapshots of periodic and non-periodic objects and dynamical systems. Each example was selected to highlight an important application area. As shown in Table~\ref{tabApplication}, a variety of  manifold-embedding techniques can be used. 

\begingroup 
\squeezetable
\begin{table*}
  \caption{\label{tabApplication}Summary of applications. For Diffusion Map see Refs.~\cite{CoifmanEtAl05,CoifmanLafon06}, Isomap Ref.~\cite{TenenbaumEtAl00}, GTM Refs.~\cite{FungEtAl08,BishopEtAl98}.}.
  \begin{ruledtabular}
    \begin{tabular}{lllll}
      Data type & Observed system & Snapshot type & Reconstruction & Manifold-embedding technique\\
      \hline
      Simulated & Adenylate kinase molecule\footnote{see Paper~I Sec.~IV~A.} & Diffraction & 3D structure & Diffusion Map \\
      & Chignolin molecule & Diffraction & 3D structure & Diffusion Map \\
      \hline
      Experimental & Superoxide dismutase-1 crystal & Diffraction & Orientation recovery & Isomap \\
      & Chaperonin molecule & Cryo-EM images & 3D structure & GTM \\
      & Pirouette & Unsorted image frames & Time series & Diffusion Map \\
      & Pas de deux & Unsorted image frames & Time series & Isomap
    \end{tabular}
  \end{ruledtabular}
\end{table*}
\endgroup

\subsection{Structure recovery from simulated diffraction snapshots of non-periodic objects at ultra-low signal}
\label{secSimulatedData}
First, we demonstrate 3D structure recovery from a collection of two million simulated diffraction snapshots of the synthetic protein chignolin (Protein Data Bank (PDB) descriptor: 1UAO, model 1) at $ 4\times 10^{-2} $ scattered photons per Shannon pixel at $1.8 \text{ \AA}$. (A Shannon pixel is of the size needed for appropriate sampling of the intensity distribution as prescribed by the Shannon-Nyquist theorem.) This scattered intensity is expected from a 500 kD protein exposed to a single pulse from an XFEL~\cite{ShneersonEtAl08,FungEtAl08}. At this signal level, Poisson (shot) noise dominates.  The ability to deal with such levels of non-additive noise was previously demonstrated only by Bayesian algorithms~\cite{FungEtAl08,LohElser09} with extremely unfavorable scaling behavior~\cite{FungEtAl08,LohElser09,SchwanderEtAl10a, GiannakisEtAl11a}, restricting the size of amenable objects to eight times the spatial resolution. 

Here, we use the symmetry-based approach described in Paper~I after modest denoising. The denoising scheme consists of two steps: (1) Convolve the snapshot pixels with a 2D Gaussian filter with a width approximately equal to that of a Shannon pixel; (2) Replace each snapshot vector by an average over its local neighbors. Depending on the SNR, a number iterations of step (2) may be needed, with a stopping criterion based on a least-squares residual determined through the first nine Laplacian eigenfunctions of the dataset (ordered in order of increasing eigenvalue). These eigenfunctions are employed in our scheme to assign an orientation to each snapshot. For details see Appendix~\ref{appendixNoisy} and Paper~I.

To estimate the accuracy of orientation recovery, we use the following measure for root-mean-square (RMS) distance between the deduced and true orientations:
\begin{equation}
  \label{eqRMSE}
  \varepsilon = \left[ \frac{ 1 }{ s ( s - 1 ) } \sum_{i\neq j}( \tilde D_{ij} - D_{ij} )^2 \right]^{1/2},
\end{equation}
where $ D_{ij} = 2 \arccos( | \tau_i \bm{ \cdot } \tau_j | )  $ and $ \tilde D_{ij} = 2 \arccos( | \tilde \tau_i \bm{ \cdot }  \tilde \tau_j | ) $ are the true and estimated 
internal distances between orientations $ i $ and $ j $, respectively, and $ \bm{ \cdot } $  is the inner product between quaternions. Moreover, to assess the influence of local averaging on the eigenfunctions employed for orientation recovery, we compute the distance $ \gamma $ of the invariant subspace $ \tilde V $ spanned by the leading nine eigenfunctions of the diffusion matrix $ \mathsf{ P }_\epsilon $ in Table~\ref{algP}  from the corresponding invariant subspace $ V $ of the noise-free diffusion matrix \footnote{$ V $ and $ \tilde V $ consist of all linear combinations of the form $ \protect\sum_{k=1}^9 c_k \protect\underline{ \psi }_k $ where $ \protect\underline{ \psi }_k$ are the first nine eigenvectors of $ \mathsf{ P }_\epsilon $ ordered in order of increasing eigenvalue.}. Note that $ \mathsf{ P }_\epsilon $ has size $ s \times s $, where $ s $ is the number of snapshots in the data set; i.e., $ \tilde V $ and $ V $ are subspaces of $ \mathbb{ R }^s $. 

Here, we employ a standard distance measure from matrix perturbation theory \cite{Stewart73}, viz.
\begin{equation}
  \label{eqDistP}
  \gamma = \lVert \tilde \Pi - \Pi \rVert_2,
\end{equation} 
 where $ \tilde \Pi $ and $ \Pi $ are orthogonal projectors from $ \mathbb{ R }^s $ to $ \tilde V $ and $ V $, respectively, and $ \lVert \cdot \rVert_2 $ denotes the spectral norm of matrices. With this definition, $ \gamma $ lies in the interval $ [ 0, 1 ] $, and may be interpreted as the sine of an angle characterizing the deviation of $ \tilde V $ from $ V $. For our purposes, Eq.~\eqref{eqDistP} is more appropriate than an error measure based on the difference between the noisy and noise-free diffusion matrices (or their generators), since the latter depends on higher eigenfunctions which are not used in our scheme.    
  
Diffraction snapshots were simulated in $ 2 \times 10^6 $ different orientations to a spatial resolution of $1.8 \text{ \AA}$ using 1 \text{\AA} photons. The orientations were sampled approximately uniformly over SO(3), as described in Ref.~\cite{LovisoloDaSilva01}. Cromer-Mann atomic scattering factors~\cite{CromerMann68} were used for the 77 non-hydrogen atoms, and the hydrogen atoms neglected. The detector pixel was the appropriate Shannon pixel~\cite{FungEtAl08}, which oversamples the scattered amplitudes by a factor of two, resulting in $ 40 \times 40 = 1600 $ Shannon detector pixels. To model shot noise, diffracted intensities were scaled so that the mean photon count (MPC) per Shannon pixel was 0.04 at $1.8 \text{ \AA}$ resolution. The quantized photon count at each pixel was obtained from a Poisson distribution by the algorithm described 
in Ref.~\cite{Knuth97}. 

\begin{figure}
  \centering\includegraphics{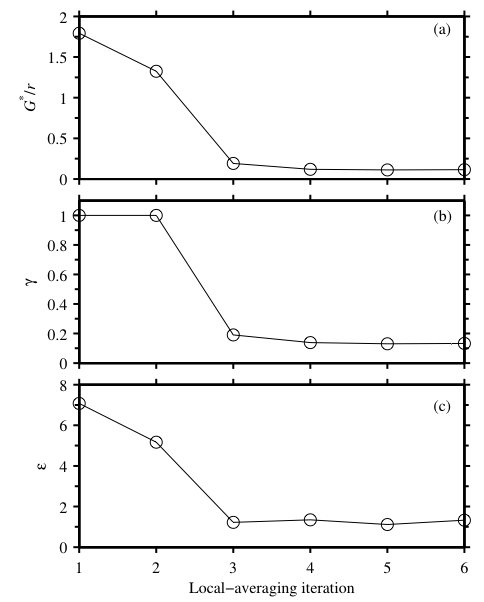}
  \caption{\label{figDenoise} (a) Least-squares residual $ \mathcal{ G }^*$, (b) invariant-subspace distance $ \gamma $, (c) RMS internal distance error $ \varepsilon $, shown as a function of the local-averaging iteration count. In Panel~(a), $ \mathcal{ G }^*$ has been normalized by the number of samples $ r = 8 \times 10^4 $ used for least-squares fitting.}
\end{figure}

With no other information, the noisy diffraction patterns were provided to the algorithm in Table~\ref{algNoisy} (width of Gaussian filter for image smoothing $ \sigma = 0.7 $; number of nearest neighbors in the sparse distance matrix $ d = 220 $; number of nearest neighbors for local averaging  $ l = 20 $; number of datapoints for least-squares fitting $ r = 8 \times 10^4 $; number of nearest neighbors for autotuning $ n =  30 $.) As illustrated in Fig.~\ref{figDenoise}, the least-squares residual $ \mathcal{ G }^* $, the subspace distance $ \gamma $, and the RMS orientation recovery error $ \varepsilon $ all decrease monotonically for the first five iterations of local averaging, but exhibit a mild increase at iteration six. At that point the algorithm was terminated in accordance with the stopping criterion described above and in Appendix~\ref{appendixNoisy}. The minimum $ \varepsilon $ value attained with this choice of parameters at iteration~5 is $ \sim 1.1 \text{ Shannon~angles} $. We measured comparable levels of orientation-recovery accuracy for various combinations of $ l $ and $ n $ parameters in the range 10--50. In all cases, we observed that small values of $ \mathcal{ G }^* $ correlate strongly with small values of $ \epsilon $, indicating that the least-squares residual provides an effective guideline for setting the parameters of the algorithm. This is particularly important, because $ \mathcal{ G }^* $ depends solely on the Laplacian eigenfunctions (see Table~\ref{algAutotuning}), and, unlike $ \varepsilon $, can be evaluated in an experimental environment where the correct orientations are not known.         

\begin{figure}[tb]
  \centering
 \includegraphics[width=\linewidth]{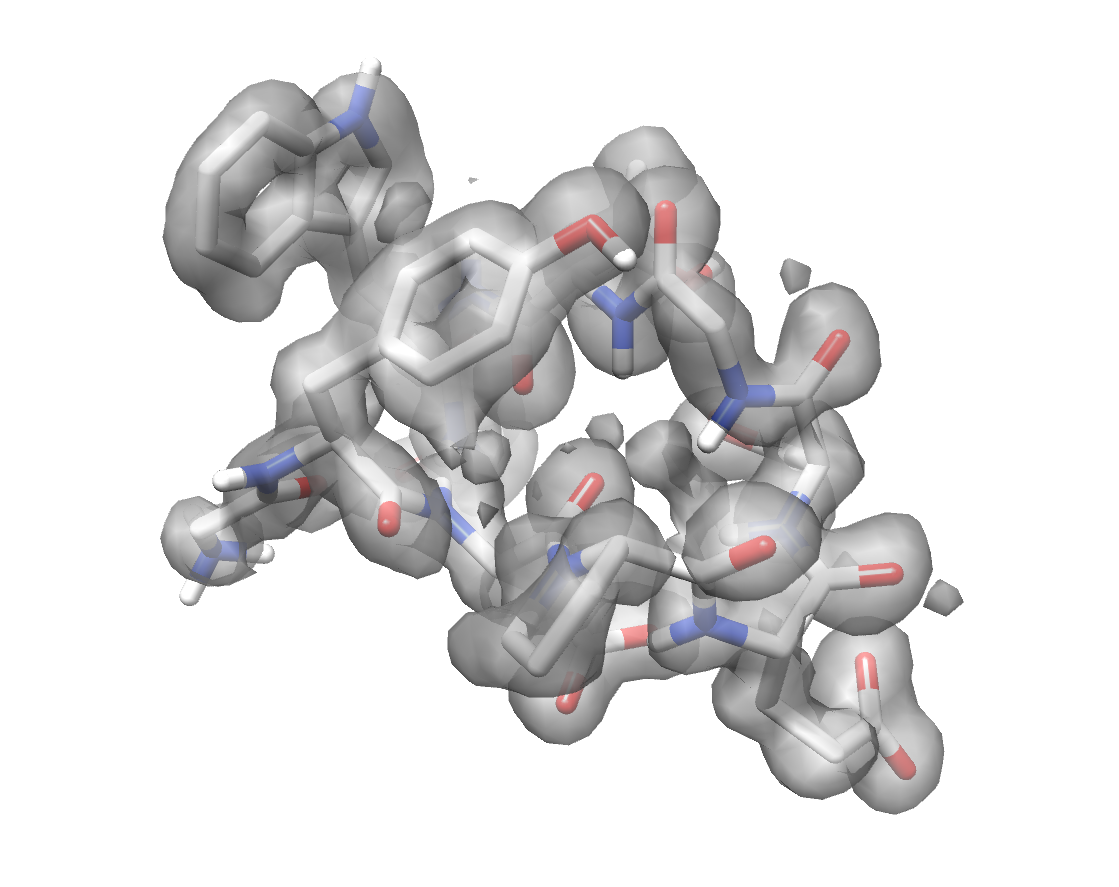}
  \caption{\label{figChignolin}Three-dimensional electron density of the synthetic protein chignolin, recovered from $ 2 \times 10^6 $  noisy diffraction patterns of unknown orientation at a mean photon count of 0.04 per pixel at $1.8 \text{ \AA}$ resolution. The ball-and-stick model represents the actual structure.}
\end{figure}

The quality of orientation recovery was further tested by inverting the reconstructed 3D diffraction volume compiled on a uniform Cartesian grid by an interpolation scheme consistent with the geometry of diffraction~\cite{SchwanderEtAl10b}. The $ R $-factor between the gridded scattering amplitudes $ \tilde F_i $  and those obtained from the Fourier transform of the recovered electron density from 
phasing, $ F_i $, was defined as 
\begin{equation}
  R = \frac{ \sum_i ( | \tilde F_i |^2 - | F_i|)^2 }{ \sum_i |F_i|^2 }.
\end{equation}  
The 3D electron density obtained by iterative phasing with the \textsc{superflip} algorithm~\cite{PalatinusChapuis07} ($ R=0.20 $) is shown in Fig.~\ref{figChignolin}. The close agreement with the known structure of chignolin clearly demonstrates sufficient alignment accuracy for reconstruction to $1.8 \text{ \AA}$ resolution. This is on par with the computationally much more expensive Bayesian approaches~\cite{FungEtAl08,LohElser09,MothsOurmazd11}. 

\subsection{Orienting diffraction patterns of crystals}
\label{appSOD}
So far we have shown that our symmetry-based approach can be used to orient diffraction patterns from single molecules to high accuracy.  We now demonstrate that this approach can also orient diffraction snapshots from crystals. This is important, because recent XFEL-based ``diffract-and-destroy'' approaches, which use femtosecond X-ray pulses to ``outrun radiation damage'', produce diffraction snapshots of nanocrystals of unknown orientation~\cite{ChapmanEtAl11}.  As a representative example, we consider a biological crystal of the enzyme superoxide dismutase-1 (SOD1, PDB designation: 1AR4) with $\sim 3\times 10^{3}$ atoms per unit cell, and thus highly complex diffraction patterns.  The key issue is whether manifolds produced by diffraction snapshots of crystals are sufficiently homogeneous (possess sufficiently homogeneous metrics) for snapshot orientations to be recovered in a straightforward manner. To demonstrate this point, we intentionally utilized snapshots spanning an orientation range of  $ 90^\circ $ so as to produce an open 1D manifold, and analyzed the dataset with the Isomap manifold-embedding method~\cite{TenenbaumEtAl00}.  In contrast to Diffusion Map, whose eigenfunctions are insensitive at the boundaries, Isomap maps a 1D open manifold to a straight line segment, and is sensitive to snapshot orientation over the entire range.   

\begin{figure}[tb]
  \centering
  \includegraphics[width=\linewidth]{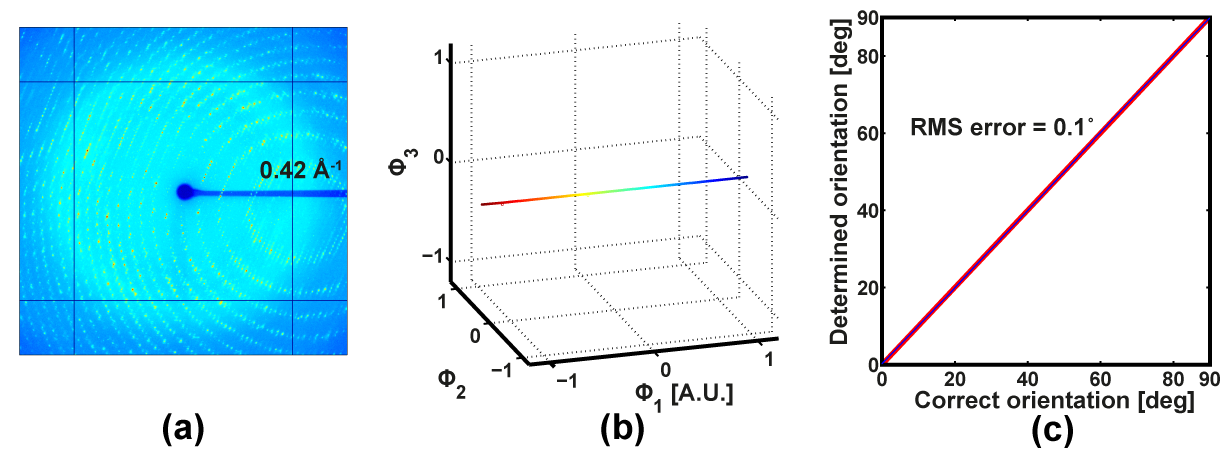}
  \caption{\label{figSOD}(a) Typical experimental diffraction pattern of superoxide dismutase-1. (b) The embedding of the geodesic distances in the space of the first three eigenvectors, with each point representing a diffraction pattern. (c) Relation between the correct and the determined orientations.}
\end{figure}

Experimental diffraction patterns of a single crystal of superoxide dismutase-1 with a mosaicity of $ 0.8^\circ $   
were obtained at the Advanced Photon Source ($\lambda = 0.98 \text{ \AA}$).
The crystal was rotated about an arbitrary axis with a step size of $ 0.5^\circ $, and 1800 diffraction patterns recorded 
over a range of $ 90^\circ $. To compensate for spurious beam intensity fluctuations, the diffraction pattern 
intensities were normalized.  Isomap was used to embed diffracted amplitudes (square-roots of intensities), 
using two nearest neighbors for calculation of geodesic distances (integrals of the metric).  As shown in 
Fig.~\ref{figSOD}, a one-dimensional and uniformly populated manifold results, with the projection on the first 
eigenvector linearly proportional to the snapshot orientation to within $ 1^\circ $, compared with the crystal 
mosaicity of $ 0.8^\circ $.  The homogeneity of this manifold (metric) establishes that, in principle, our 
symmetry-based approach can be used to treat crystalline objects in the same way as non-periodic single
particles, provided, of course, object symmetry is appropriately incorporated. 

\subsection{Structure recovery from experimental cryo-electron micrographs}
\label{appCryo}
A well-studied application of graph-theoretic techniques concerns the 3D reconstruction of faint 
biological objects by single-particle cryo-EM without orientational information. 
In cryo-EM, the resolution is strongly degraded by radiation damage. 
As such, the lowest acceptable exposure to electrons and thus SNR must be employed.  As described in Sec.~\ref{secPreviousWork}, this has 
proved a fertile ground for testing new algorithms.
By recourse to specific properties of cryo-EM images, impressive results have been obtained, primarily with simulated data.  Beyond noise, however, 
reconstruction by cryo-EM must contend with a range of key issues, chief among them the loss of 
information due to zero-crossings in the transfer function of the microscope and thus partial loss of 
information in any single snapshot. The exact position of the zero-crossings depends sensitively on microscope defocus.  This 
offers a means to recoup some of the lost information by insuring that the dataset includes micrographs 
obtained over a range of defocus values, each with a different set of zero-crossings in the transfer function.
The key point is this:  the object structure cannot be recovered in full detail from a single defocus, even if the imaging parameters were known exactly.
Thus, for a reconstruction algorithm to be of practical use, it must deal with the effect of defocus variations --- a test rarely passed by 
new algorithms.  Here, we demonstrate structure recovery from experimental cryo-EM images of the biological molecule chaperonin. 
Specifically, we incorporate the effect of defocus, use 
the symmetry-based homogeneity of the manifold metric to deduce orientations, and thence recover the 3D 
object structure.  This demonstration is mitigated by two factors:  (1) in order to expedite the calculations, 
snapshots with only one orientational degree of freedom were selected from a set presorted by a standard 
orientation algorithm; and (2) to demonstrate structure recovery at ultra-low signal --- far below what is 
normally used --- experimental snapshots were preprocessed  to simulate such low signal levels. 

Randomly oriented single-particle cryo-EM images images of the wild-type group~II chaperonin in \emph{methanococcus maripaludis} (Mm-cpn), obtained with a mean incident electron count (MEC) of $ 20/\text{\AA}^2 $ (equivalent to 135 electrons per $2.6 \text{ \AA}$-square snapshot pixel) were kindly provided by Chiu \emph{et al}.\ \cite{ZhangEtAl10}.  Each snapshot consisted of $ 96 \times 96 $ pixels. 
A set of 413 side-view snapshots was  selected from 5000 images, whose orientations had been previously determined by the
\textsc{eman} program~\cite{LudtkeEtAl99}, resulting in a dataset with a single orientational degree of freedom about the object symmetry axis.  

To investigate the performance of our method at lower dose, a second data set was produced by applying an additional Poisson process to the raw experimental images. The method is based on an approximation valid for low-contrast images with Poisson noise and sufficiently large MEC. The substitution $ I \mapsto I' = \Pois( I^{1/2} ) $ transforms a signal $ I $ to a signal $ I' $, with mean $ \text{MEC}' = \text{MEC}^{1/2} $ and variance $ \var( I' ) = \var( I ) / 4 $. Simulations verified the accurate validity of this approach at MEC = 100, compared with an MEC per snapshot pixel of 135 for our experimental images.  Twenty noisy versions of each image were thus generated to form a data set of 8260 images with an effective MEC of 1.7 per $\text{\AA}^2$ .  

Since neither the noise-free signal nor the noise variance was known for our experimental cryo-EM images, 
a method developed by Frank~\cite{Frank06} was used to estimate the SNR directly from the experimental data. 
This determines the SNR from the cross-correlation coefficient  $ C_{ij} $  between two images in the same 
orientational class using the definition: 
\begin{equation}
\label{eqSNR_I}
  \text{SNR} = 10 \log_{10} \mmean( C_{ij} / ( 1 - C_{ij} ) ),
\end{equation}
where the mean is taken over all classes and all images within each class. Provided two images 
represent different realizations of noise from an identical object in the same orientation, the above estimate 
for SNR agrees with the standard definition $ \text{SNR}=10\log_{10} \frac{\var(signal)}{\var(noise)} $~\cite{FrankAli75}.
With the classification obtained from \textsc{eman}~\cite{LudtkeEtAl99}, and the assumption that class members differ only in noise, we estimate a SNR of $ -6\text{ dB} $ for the raw experimental snapshots (MEC:~$ 20/\text{\AA}^2 $) and a SNR of $ -16\text{ dB} $ for the preprocessed experimental snapshots (MEC:~$ 1.7/\text{\AA}^2 $). 

As described above, the defocus value and hence transfer function of the electron microscope vary from 
snapshot to snapshot. To analyze such cryo-EM data, we implemented a modified version of the 
manifold-embedding algorithm Generative Topographic Mapping (GTM)~\cite{BishopEtAl98,Svensen98} to explicitly incorporate 
the effect of the microscope transfer function.  GTM defines a manifold in data space by partitioning the 
noisy dataset into a number of Gaussians each centered around a point (node) on the manifold.  The 
partitioning is based on a nonlinear mapping of a latent space, in this case the space of rotations.  GTM is 
thus, in essence, a manifold-embedding technique, with the symmetries of scattering manifested in the 
homogeneity of the data manifold, as described in Paper~I.  However, the generative capability of GTM allows one to construct an image (in essence a model snapshot) at each node on the data manifold.  In our
approach, this model image extracted from the data corresponds to the aberration-free projected potential of the object.  In order to 
assign an experimental snapshot to a model image, its distance from the model is calculated after 
convolving the model with the transfer function of the microscope at the defocus corresponding to that of 
the experimental snapshot.  This convolution proceeds efficiently as multiplication in Fourier space, and is 
not computationally expensive. 
A similar approach based on more efficient manifold-embedding techniques will be published elsewhere.

The GTM-based approached was first validated with simulated cryo-EM images of chaperonin over a 
typical experimental defocus range  of $10,000 \text{ \AA}$ to $30,000 \text{ \AA}$ (underfocus).  The orientations were successfully recovered 
to within~$ 1^\circ $. To reconstruct 3D density maps from experimental snapshots, model aberration-free 
projected potentials were generated (lifted) at 16 equally-spaced nodes of the data manifold produced by 
experimental images replicated according to the 8-fold object symmetry. 3D density maps were then 
reconstructed tomographically using the back-projection algorithm \textsc{bg~cg} of the \textsc{spider} software package \cite{FrankEtAl96}. 
For comparison, a simulated density map was obtained from 2D snapshots using the known chaperonin 
atomic coordinates (PDB identifier: 3LOS) under the following imaging conditions: spherical aberration $ C_s = 4.1 \text{ mm} $; 
 defocus $ \Delta f=\text{24,000 \AA} $ (underfocus); electron energy $ E = 300 \text{ keV} $; damping envelope 
parameter $ B=50 \text{ \AA}^2$; images phase-flipped.  The resulting 3D density map was passed through a $ 5 $ \text{\AA} 
Gaussian filter. 

\begin{figure}[tb]
  \centering
  \includegraphics[width=\linewidth]{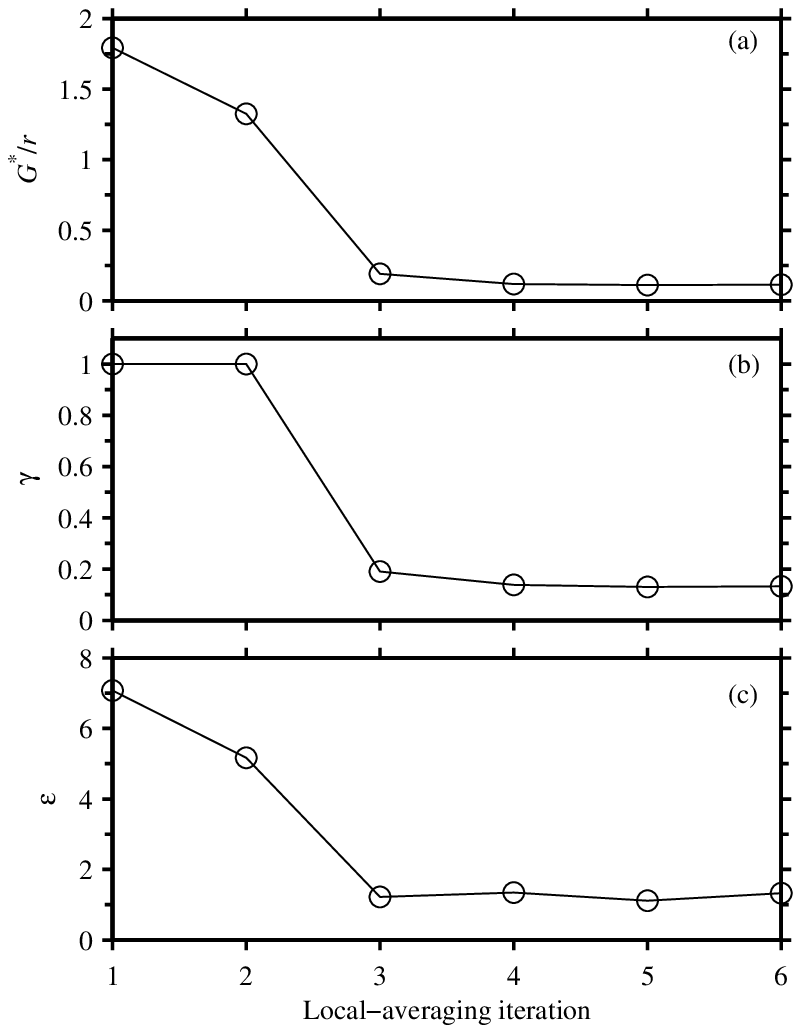}
  \caption{\label{figChaperonin}(a) Experimental cryo-electron micrograph of a chaperonin molecule at a mean electron count of $20/\text{\AA}^2$ ($ \text{SNR} = -6\text{ dB}$).  (b) Image obtained by averaging the members of an orientation class. (c) Image generated (lifted) from the data manifold.  (d) Experimental micrograph of chaperonin molecule processed to reflect a mean electron count of $1.7/\text{\AA}^2$ ($\text{SNR} = -16\text{ dB}$).  (e) Image lifted from the data manifold.  (f) 3D reconstruction with simulated images. (g) 3D reconstruction with lifted images at a snapshot dose of $20/\text{\AA}^2$.  (h) 3D reconstruction with lifted images at a snapshot dose of $1.7/\text{\text{\AA}}^2$.}
\end{figure}

Fig.~\ref{figChaperonin}(a) shows a typical experimental snapshot, Fig.~\ref{figChaperonin}(b) the average of the micrographs assigned to an 
orientation class by the cryo-EM reconstruction software package \textsc{eman}~\cite{LudtkeEtAl99}, and Fig.~\ref{figChaperonin}(c) the snapshots 
oriented by manifold embedding and reconstructed (lifted) from the manifold. (For a movie of the 
reconstructed tilt series see EPAPS Movie~1). Note that the manifold is able to 
generate missing images by interpolation. The improved quality of the manifold-generated snapshots 
compared to the class averages offers the possibility to reconstruct at significantly reduced dose. Fig.~\ref{figChaperonin}(d) is 
an experimental snapshot preprocessed to approximate snapshots expected from a single chaperonin 
molecule at a dose $12\times$ lower than commonly used~\cite{ZhangEtAl10} ($\text{SNR}\sim-16 \text{ dB}$, i.e., 10 dB below a typical dose). 
Fig.~\ref{figChaperonin}(e) is the snapshot lifted from the manifold after orienting an ensemble of 8000 different raw 
snapshots by manifold embedding. It is clear from this image, the corresponding tilt series (EPAPS Movie~2), and the 3D reconstructions of Fig.~\ref{figChaperonin}(f--h) that snapshots can be successfully oriented by manifold 
embedding to produce 3D models, even at $ 12\times $ lower signal than in use today. 
Note that  images at this dose could not be 
oriented by standard cryo-EM approaches~\cite{FrankEtAl96}, even when accurately centered prior to analysis, as was 
performed here. In contrast, our orientation recovery results were similar to those obtained at an MEC of $20/\text{\AA}^2$, indicating that the effect of lower signal levels can be compensated by increasing the number of snapshots. 
 
Our results thus offer the tantalizing possibility of reducing the snapshot dose in 3D reconstruction techniques using 
ionizing radiation, in some cases by at least an order of magnitude. This would significantly mitigate the limits set by 
radiation damage. As a benchmark, the essentially unfulfilled promise of the costly transition of cryo-EM to liquid He 
temperatures was aimed at improving dose tolerance by a factor of two. The superior signal extraction 
capability offered by manifold mapping could also be used to obtain images at smaller defocus values in 
order to reconstruct the object to higher resolution. 

\subsection{Time-series (movies) from ultra-low-signal random-sequence snapshots}
\label{appPirouettePasDeDeux}
Our knowledge of the precise time at which an experimental snapshot of a dynamic system was obtained is 
corrupted by inevitable uncertainties, which can substantially exceed the intrinsic time 
resolution of the observation technique.  Modern pump-probe experiments, for example, can now be 
performed with pulses as short as a few femtoseconds, but their time-resolution is often determined by 
timing jitter, which can be up to two orders of magnitude larger~\cite{Glownia:10,Cryan10}.
When the state of a system under observation is not synchronized with the observation 
windows, a sequence of snapshots can represent random sightings of the system during its evolution.  It is 
thus important to develop means for deducing ``time stamps'' directly from the data, either to reduce jitter-induced uncertainty, 
or to order a sequence of snapshots according to the intrinsic evolution of the system 
under observation.  Here, we demonstrate this capability at SNRs as low as $-21\text{ dB}$.  Specifically, we show 
that: (1) Randomized movie sequences can be time-ordered, even when the signal is extremely low; and (2) 
Frames generated (lifted) from the manifold produce movies of superior quality. 

Movies of a pirouette and a pas de deux were downloaded from the web.  These represent 
optical snapshots of a conformationally rigid body in rotations, and the evolution of two flexible bodies in 
interaction, respectively. In order to reduce the SNR, a constant background was added, and shot noise incorporated at 
each pixel depending on its intensity value, as described in Ref.~\cite{Knuth97}.  For the pirouette, a sequence of 16~turns 
consisting of 268 frames ($ 210\times160 $ pixels each) was replicated 132 times, a background $ 5\times $ the mean intensity added, and shot noise incorporated to produce an effective mean photon count per pixel of 0.08 and a SNR of $-21\text{ dB}$ (see Eq.~\ref{eqSNR_I}).  For the pas de deux, a sequence of 870 frames ($ 265\times305 $ pixels each) was replicated 12 times with an added background of twice the mean intensity, and shot noise incorporated to produce a mean photon count of 0.8 and a SNR of $ -11\text{ dB} $. For both movies, camera motion was corrected by reference to a stationary marker.  

Each random sequence was ordered by a suitable manifold-embedding technique (Diffusion Map or 
Isomap).  Using the generative property of GTM, images were then lifted from the manifold.  As described 
in Ref.~\cite{SchwanderEtAl10a} and demonstrated in Fig.~\ref{figChaperonin}(c,e), this procedure uses the information content 
of the entire dataset to generate each snapshot, producing images of significantly higher quality than 
possible by traditional classifying and averaging techniques.
It is also more robust against non-uniform sampling and jitter. Table~\ref{algLifting} summarizes the lifting procedure.

\begingroup 
\squeezetable
\begin{table}
  \caption{\label{algLifting}Manifold-lifting algorithm based on GTM}
  \begin{ruledtabular}
    \begin{tabular}{c}
      \begin{minipage}{\linewidth}
        \begin{algorithmic}[1]
          \Statex \textbf{Inputs:}
          \Statexi  Noisy snapshots $ \mathcal{ M }_I = \{ \underline{I}_1,\ldots,\underline{I}_s \} $
          \Statexi Estimated quaternions  $ \mathcal{ T } = \{ \tau_1, \ldots, \tau_s \} $
          \Statexi number of nodes $ K $ 
          \Statexi number of basis functions $ M $
          \Statexi basis function width $ \sigma $
          \Statex
          \Statex \textbf{Outputs:}
          \Statexi Manifold-lifted images $ \mathcal{ M } = \{ \underline{a}_1, \ldots, \underline{a}_s \} $
          \Statex
           \State Generate the grid of latent points $ \{ x_1, \ldots, x_K \} $.  
           \State Generate the grid of basis function centers $ \{ \mu_1, \ldots, \mu_M \} $. 
           \State Compute the matrix of basis function activations $ \Phi $ such that
           \begin{displaymath}
             \Phi_{m}( x ) = \exp( - ( x - \mu_m )^2/ 2 \sigma^2 ).
           \end{displaymath}
           \State Initialize a set of weights $ W $ using principal component analysis.
           \State Initialize inverse Gaussian noise variances $\alpha $ and $\beta$. 
           \State Compute a set of responsibilities $ R $ that assigns each snapshot to a node from the results of manifold embedding. 
           \State Compute the diagonal matrix $ G $ using $ R $, where $ G_{kk} = \sum_{i=1}^s R_{ki} $.
           \Repeat
           \State $ W \leftarrow ( \Phi^\mathrm{ T } G \Phi + \lambda I )^{-1} \Phi^\mathrm{ T } R \mathcal{ M }_I $, where the regularization parameter $ \lambda $ may be zero. 
           \State Compute  $ \Delta $, where $ \Delta_{kn} = \lVert \underline{ I }_n - \Phi_k W \rVert^2 $.
           \State Compute $ \gamma $ from $ \lambda $ and $ \alpha $.
           \State Update $ \alpha $ and $ \beta $ using $ \gamma $, $ R $ and $ \Delta $.
           \Until{convergence of $ W $}
           \State $ \mathcal{ M } \leftarrow \Phi W  $
           \State \Return $ \mathcal{ M } $
         \end{algorithmic}
       \end{minipage}
     \end{tabular}
   \end{ruledtabular}
\end{table}
\endgroup

\begin{figure}[tp]
  \centering
  \includegraphics[width=\linewidth]{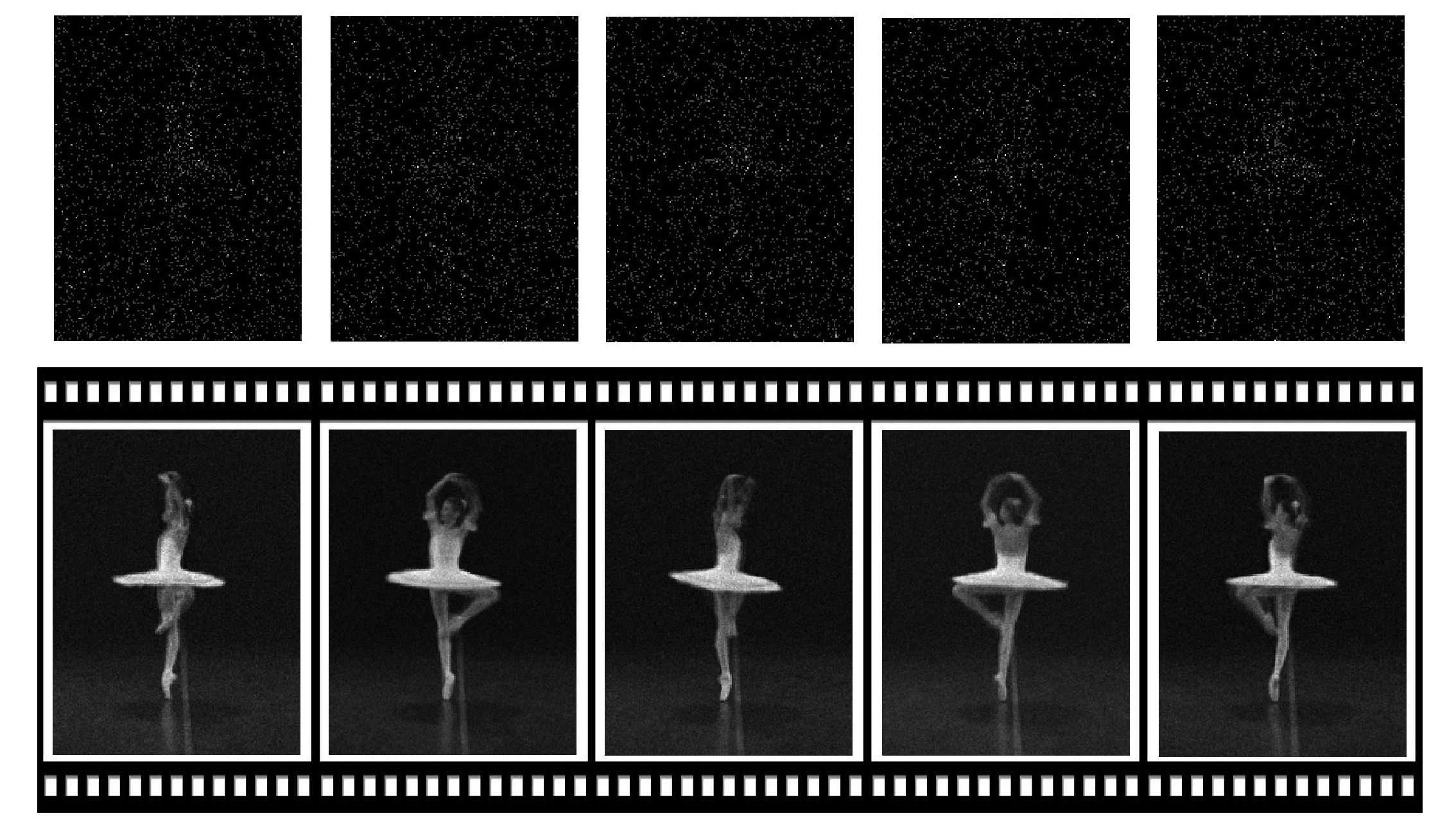}
  \caption{\label{figPirouette}Top row: First five frames of 35,000 randomly-sequenced snapshots of a pirouette preprocessed to reflect a mean photon count of 0.08 per pixel with added background and shot noise, corresponding to a signal-to-noise ratio of $ -21 $ dB. Bottom row: Five evenly-spaced images extracted from the Diffusion Map manifold. (See also EPAPS Movie~3.)}
\end{figure}

\begin{figure}[tp]
  \centering
  \includegraphics[width=\linewidth]{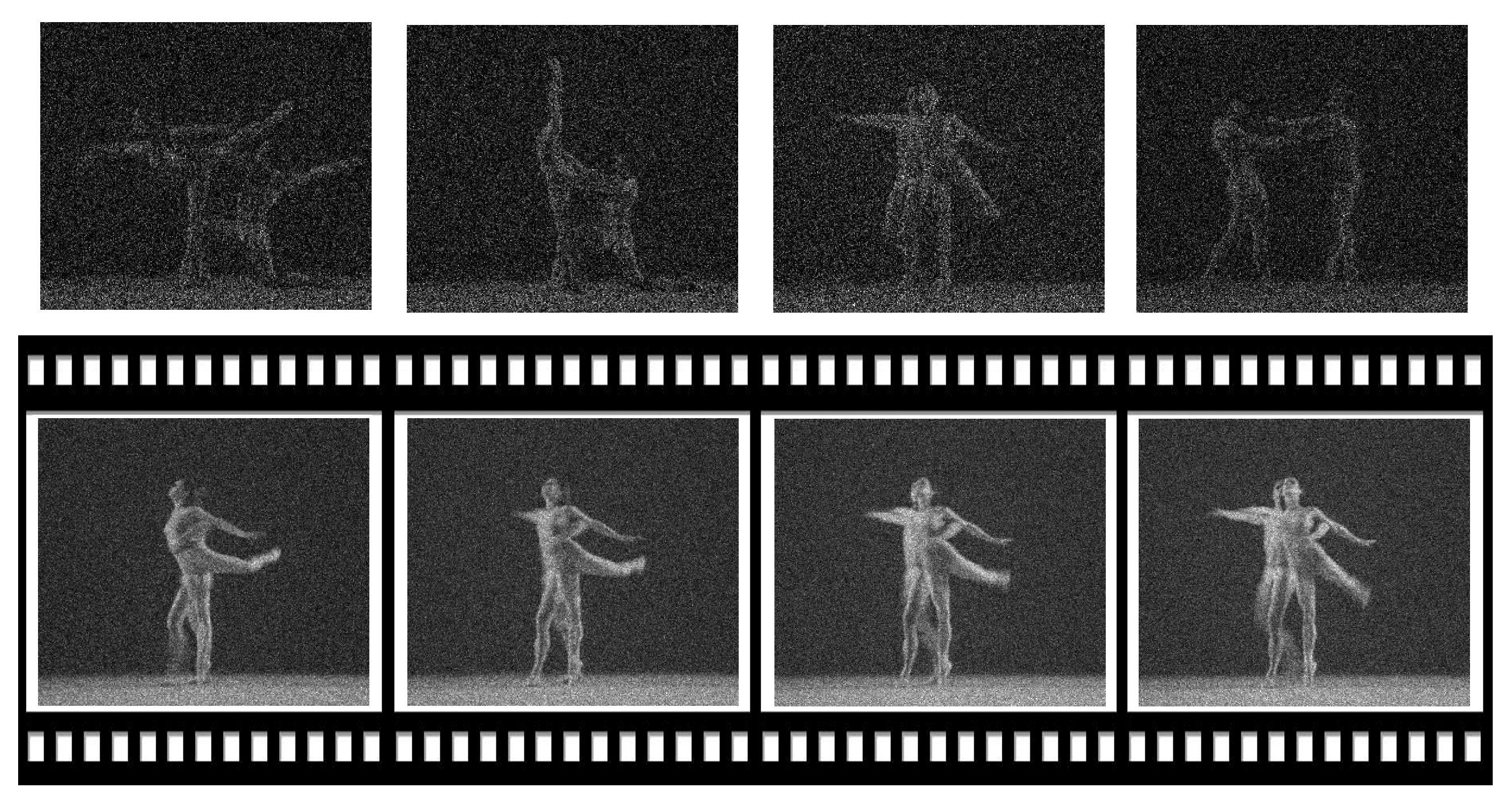}
  \caption{\label{figPasDeDeux}Top row: First four frames of 10,000 randomly-sequenced snapshots of a pas de deux preprocessed to reflect a mean photon count of 0.8 per pixel with added background and shot noise, corresponding to a signal-to-noise ratio of $ -11 $ dB. Bottom row: Four evenly-spaced images extracted from the Isomap manifold.  (See also EPAPS Movie~4.)}
\end{figure}

For the pirouette, Diffusion Map was used to recover the object orientation in each frame during the dance, and hence the sequence order (number of nearest neighbors in the sparse distance matrix $ d = 5896 $; Gaussian kernel bandwidth $ \epsilon = 200 $.)  Snapshots were then lifted from the manifold (number of nodes $ K = 28 $; number of basis functions $ M = 14 $, basis function width $ \sigma = 2 $.) Reconstructed images are shown in Fig.~\ref{figPirouette}
together with a sequence of unsorted, unprocessed snapshots.  The movie is available as EPAPS Movie~3. 
The randomized pas de deux sequence was ordered with Isomap (number of nearest neighbors $ d = 33 $.) To compile the movie, 
images were lifted from an ordered sequence of 870 points (nodes), corresponding to uniform 
sampling on the Isomap manifold. Reconstructed images are shown in Fig.~\ref{figPasDeDeux} together with a sequence of unsorted snapshots.
The movie is available as EPAPS Movie~4.  

Figs.~\ref{figPirouette}~and~\ref{figPasDeDeux}, and the associated movies clearly show that our approach is able to determine the correct frame sequence and generate high quality snapshots at signal levels as low as 0.08 
photon/pixel for the pirouette (modulo one revolution), and 0.8 photon/pixel for the pas de deux, both with 
added background and non-additive noise corresponding to signal-to-noise ratios in the range $-11 $ to $-21\text{ dB} $. These examples demonstrate the capability to determine the time evolution of systems from unsorted 
random sightings at extremely low signal.  They also highlight the potential to correct timing jitter in 
pump-probe experiments, and reconstruct the evolution of dynamic systems from random sightings of members of a 
heterogeneous set, each at a different stage of its evolution. These possibilities will be described 
in detail elsewhere. The general implications for signal extraction and image processing are clear.  

\section{Conclusions}
\label{secConclusions}
We have shown that manifold mapping, as described in Paper~I, augmented with modest noise reduction 
measures, is able to extract structural and timing information from simulated and experimental snapshots at 
extremely low signal.  The ability to orient simulated and experimental diffraction and image snapshots 
confirms the accessibility of the homogeneous manifold expected from our theoretical framework for a wide 
range of objects and imaging scenarios, including crystalline samples.  The capability to recover 3D 
structure at extremely low signal is on par with the more expensive Bayesian approaches, but offers greater 
reach in terms of sample size and resolution, as demonstrated in Paper~I.  The noise-robustness of our approach 
substantially exceeds what has been demonstrated with comparable graph-theoretic approaches
without restrictive, application-specific assumptions.  The manifold itself offers a powerful route to image 
reconstruction at low signal, because snapshots reconstructed from the manifold achieve higher signal-to-noise ratios than 
possible by traditional approaches based on classification and averaging.  
Taken together, these offer a physically-based, computationally efficient, noise-robust route to analyzing the large and 
varied datasets generated by existing and emerging structure recovery methods.  In the longer term, it
should be possible to use these approaches to recover or improve timing information in pump-probe 
experiments, and construct 3D movies (4D maps) from random sightings of members of structurally 
heterogeneous and dynamically evolving ensembles. 


%
%

%

\begin{acknowledgments}
We are grateful for experimental data and advice to W.\ Chiu and J.\ Zhang (cryo-EM images) and M.\ 
Schmidt (crystallographic data), and acknowledge discussions with G.\ N.\ Phillips, Jr., R.\ Rosner, W.\ 
Schr\" oter, and members of the UWM Physics Department. We are grateful to J.\ Frank and H.\ Y.\ Liao for 
assistance with 3D reconstruction by back-projection. One of us (DG) is grateful to the Random Shapes 
Program held in 2007 at the Institute for Pure \& Applied Mathematics. This work was partially supported 
by the U.S. Department of Energy Office of Science (SC-22, BES) award \#DE-SC0002164 and a UWM 
Research Growth Initiative award. 
\end{acknowledgments}

\appendix
  
\section{Treatment of noise}
\label{appendixNoisy}

In the manifold picture, noise can be described as a perturbation of the noise-free manifold $\mathcal{ M }  = \{ \underline{  a }_1, \underline{ a  }_2, \ldots, \underline{ a  }_{s} \} $, where $ \underline{  a }_i $ is a snapshot vector of measured pixel amplitudes, viz. 
\begin{subequations}
  \begin{gather}
    \label{eqSnapshotNoisy}
    \underline{ a }_i \mapsto \underline{ \tilde a }_i = \kappa \underline{ a }_i + \delta \underline{ a }_i, \\
    \delta \underline{ a }_i = ( \delta a_{i1}, \delta a_{i2}, \ldots, \delta a_{in} )^\mathrm{T}, \quad \delta a_{ij} = I_{ij}^{1/2} - \kappa a_{ij}\;.  
  \end{gather}
\end{subequations}
This causes the observed, noisy, dataset   
\begin{equation}
  \label{eqDataSetANoisy}
  \tilde{ \mathcal{ M } } = \{ \underline{ \tilde{ a } }_1, \underline{ \tilde{ a } }_2, \ldots, \underline{ \tilde{ a } }_{s} \} 
\end{equation}
not to lie exactly on the manifold $ \mathcal{M} $ (up to a global scaling by $ \kappa $). One would expect that if the perturbation norm $ \lVert \delta \underline{ a }_i \rVert $ becomes comparable to the kernel bandwidth $ \epsilon^{1/2} $, the computed eigenfunctions $ \underline{ \psi }_k $ are distorted to the point that the embedded manifold no longer has the topology of SO(3) \cite{CoifmanLafon06,BalasubramanianSchwartz02}. Indeed, as illustrated in Fig.~\ref{figDenoise} direct application of the algorithm for noise-free data (see Paper~I Table~III) to a noisy dataset $ \tilde{ \mathcal{ M } } $ at an $ \text{MPC} = \Ord( 10^{-2} ) $ results in poor accuracy. In order to be practically useful, the noise-free orientation-recovery scheme must be augmented by a suitable denoising method.

Conceptually, we denoise the data in three steps: (1) Low-pass filtering of each snapshot by convolution of pixel intensities with a 2D Gaussian; (2) Variance-stabilizing transformation (VST); and (3) Local averaging over nearest neighbors in data space prior to embedding.  In practice, we combine (1) and (2), known to be effective for shot noise~\cite{ZhangEtAl07,ZhangEtAl08,Guan09}, into a single step, and follow the iterative procedure described below. 

\subsection{Low-pass Gaussian filtering and variance-stabilizing transformation (VST)}
Convolution with a low-pass Gaussian filter of bandwidth $ \sigma $ is represented by:
\begin{subequations}
  \label{eqGaussianFiltering}
  \begin{equation}
    \underline{ I } \mapsto ( I \ast H_\sigma )( \vec{ r } ) = \int d \vec{ r }' \, I( \vec{ r }' ) H_\sigma( \vec{ r } - \vec{ r }' ), 
  \end{equation}
  with
  \begin{equation} 
    H_\sigma( \vec{ r } ) = \exp( - \lVert \vec{ r } \rVert^2 / 2 \sigma^2 ) / ( 2 \pi \sigma^2 )^{1/2}.
  \end{equation}
\end{subequations}
VST, proposed by Guan~\cite{Guan09} for low-intensity data, can be written as:
\begin{equation}
  \label{eqVST1}
  \underline{ I }  \mapsto \underline{ I }^{1/2} + ( \underline{ I } + 1 )^{1/2}.
\end{equation}
$ I( \vec{ r } ) $ denotes the (discretely sampled) intensity pattern on the detector plane obtained by ``unpacking'' the column vector of intensities $ \underline{ I } $. 

Eqs.~\eqref{eqGaussianFiltering} and~\eqref{eqVST1} are combined into a single operation:
\begin{equation}
  \label{eqVST}
  \VST( I; \sigma ) = ( I \ast H_\sigma )^{1/2} + [ ( I \ast H_\sigma ) + 1 ]^{1/2}.
\end{equation}
Given an intensity-pattern dataset $ \mathcal{ M }_I $ consisting of $ s $ samples and an index matrix of nearest neighbors $ \mathsf{ N } $ of dimensions $ s \times l $, we introduce a combined VST and aggregation operation taking $ \mathcal{ M }_I $ to a dataset $ \tilde{ \mathcal{ M } }= \VSTL( \mathcal{ M }_I; \sigma, \mathsf{ N } ) $ such that
\begin{equation}
  \label{eqVSTL}
  \tilde{ \mathcal{ M } } = \{ \underline{ \tilde{ a } }_1, \underline{ \tilde{ a } }_2, \ldots, \underline{ \tilde{ a } }_{s} \}, \quad \underline{ \tilde{ a } }_i = \VST\left( \sum_{k=1}^{l} \underline{ I }_{N_{ik}}; \sigma \right) .
\end{equation} 
Noise robustness can be further enhanced by a so-called self-tuning Gaussian kernel introduced by Zelnik-Manor and Perona \cite{ZelnikManorPerona04}. Here, instead of the isotropic Gaussian kernel $ \mathcal{ K }( \underline{ a }_i, \underline{ a }_j ) = \exp( - \lVert \underline{ a }_i - \underline{ a }_j \rVert^2 / \epsilon )  $ of Eq.~(B3)  in Paper~I,
one uses an anisotropic Gaussian kernel with local scaling parameters, $ \epsilon_i $, given by
\begin{equation}
  \label{eqAutotuningKernel}
  \mathcal{ K }( \underline{ a }_i, \underline{ a }_j ) = \exp( - \lVert \underline{ a }_i - \underline{ a }_j \rVert^2 / ( \epsilon_i \epsilon_j )^{1/2} ).
\end{equation}
A canonical choice for the scaling parameters, which we adopt throughout, is $ \epsilon_i = \lVert \underline{ a }_i - \underline{ a }_{N_{il}} \rVert^2 $, where, as usual, $ N_{il} $ denotes the index of the $ l $-th nearest neighbor of datapoint $ i $.

\subsection{Iterative local averaging}
If the true nearest neighbors of a point in data space are known, and noise produces no systematic bias, local averaging approaches the true manifold. To see this, let $ \varepsilon $ be an error tolerance  in data space, and consider a reference orientation $ \mathsf{ R } $ with corresponding noise-free snapshot $ \underline{ a } $. For any $ \varepsilon > 0 $ it is possible to find a ball $ B_\varepsilon $ in data space centered at $ \underline{ a } $, such that for any countable set $ \{ \underline{ a }_1, \ldots, \underline{ a_l } \} $ of noise-free snapshots lying in $ B_\varepsilon $ the mean, $ \bar{ \underline{ a } } = \sum_{i=1}^l \underline{ a }_i / l $,  has error $ \lVert \bar{ \underline{ a }  } - \underline{ a } \rVert < \varepsilon $.  In the presence of noise, the $ \underline{ a }_i $ are replaced by the random variables in Eq.~\eqref{eqSnapshotNoisy}; i.e., $ \underline{ a }_i \mapsto \tilde{ \underline{ a } }_i $, where $ \tilde{ \underline{ a } }_i $ are statistically independent, have expectation value $ \kappa^{1/2} \underline{ a }_i $ proportional to  $  \underline{ a }_i $, and finite variance $ \Delta \underline{ a }_i^2 $. Moreover, the sample mean within $ B_\varepsilon $ becomes a random variable $ \hat{ \underline{ a } } = \sum_{i=1}^l \tilde{ \underline{ a } }_i / l $ with expectation value $ \kappa^{1/2} \bar{ \underline{ a } } $ and variance $ \Delta \underline{ a }^ 2 = \sum_{i=1}^l \Delta \underline{ a }_i^2 / l $. By the law of large numbers, in the limit of an infinite data set with infinite snapshots in $ B_\varepsilon $ (i.e., $ l \to \infty $), $ \hat{ \underline{ a } }_i $ is equal to $ \bar{ \underline{ a } }_i $ (up to an unimportant proportionality constant)  with probability one. Thus, recovery of the data manifold with error $ \varepsilon $ is possible almost surely. 

In practice, noise corrupts the local neighborhood relations, and without \emph{a priori} information, it is not possible to identify which of the snapshots in a noisy data set are associated with the ball $ B_\varepsilon $ of the underlying noise-free system.  Therefore, it cannot be guaranteed that local averaging leads to the correct manifold.  We therefore exploit our knowledge of the natural eigenfunctions of scattering manifolds to monitor the effect of local averaging, and terminate the procedure before substantial deviations have occurred.  

Specifically, we follow the algorithm described in Table~\ref{algNoisy}. First, we apply the $ \VST $ operation~\eqref{eqVST} to the intensity data $ \{ \underline{ I }_i \} = \mathcal{ M }_I $, setting the filter width $ \sigma $ to a relatively small value (e.g., in Sec.~\ref{secSimulatedData}, $ \sigma $ is set to $7/10$ of a pixel width). The autotuning version of the orientation-recovery method (Table~\ref{algAutotuning}) is executed using the $ \VST $-filtered intensities as input data. The nearest-neighbor indices $ \mathsf{ N }_0 $ obtained in the course of the calculation of the sparse distance matrix then become our initial estimate for the true nearest-neighbor indices. We also record the residual of the nonlinear least-squares output, $ \mathcal{ G }^*_{0} $, and choose a value $ l \leq d $ for the number of nearest neighbors for local averaging. 

\begingroup 
\squeezetable
\begin{table}
  \caption{\label{algNoisy}Orientation-recovery for noisy snapshots}
  \begin{ruledtabular}
    \begin{tabular}{c}
      \begin{minipage}{\linewidth}
        \begin{algorithmic}[1]
          \Statex \textbf{Inputs:}
          \Statexi Noisy snapshots $ \mathcal{ M }_I = \{ \underline{ I }_1, \ldots, \underline{ I }_s \} $
          \Statexi Number of retained nearest neighbors $ d $
          \Statexi Number of nearest neighbors for local averaging
          \Statexi Number of datapoints in the least-squares fit, $ r $
          \Statexi Number of nearest neighbors for autotuning, $ n $
          \Statexi Gaussian filter bandwidth $ \sigma $
          \Statex
          \Statex \textbf{Outputs:}
          \Statexi Estimated quaternions $ \mathcal{ T } = \{ \tau_1, \ldots, \tau_s \} $
          \Statexi Estimated nearest-neighbor index matrix $ \mathsf{ N } $  
          \Statexi Least-squares residual $\mathcal{ G }^* $
          \Statex
          \For{$i=1, \ldots, s$} 
          \State $ \underline{ \tilde a }_i \leftarrow \VST( \underline{ I }_i; \sigma ) $ 
          \EndFor
          \State $ \tilde{ \mathcal{ M } }_0 \leftarrow \left\{ \underline{ \tilde a }_i \right\} $ \Comment{initial iterate for Diffusion Map input data}
          \State Execute the algorithm in Table~\ref{algAutotuning} with input data $ \tilde{\mathcal{ M }}_{0} $; store the returned nearest-neighbor index matrix as $ \mathsf{ N }_{0} $ and  the least squares residual as $ \mathcal{ G }^*_{0} $.
          \State $ i \leftarrow 1 $ \Comment{initialize iteration counter.}
          \State $ \ifterm \leftarrow \text{false} $ \Comment{initialize termination flag.}
          \While{$ \ifterm \equiv \text{false} $}
          \State $ \tilde{\mathcal{ M }}_{i} \leftarrow \VSTL( \mathcal{ M }_I; \sigma, \mathsf{ N }_{i-1} ) $ \Comment{current iterate for Diffusion Map input-data} 
          \State Execute the algorithm in Table~\ref{algAutotuning} with input data $ \tilde{\mathcal{M}}_{i} $; store the outputs as $ \mathcal{ T }_{i} $, $ \mathsf{ N }_{i} $, and $ \mathcal{ G }^*_{i} $.
          \State $ \ifterm \leftarrow \mathcal{ G }^*_{i} > \mathcal{ G }^*_{i-1} $ \Comment{terminate if the residual has increased.}
          \If{$ \ifterm \equiv \text{false}$}
          \State $ i \leftarrow i + 1 $ \Comment{increment iteration counter.}
          \EndIf
          \EndWhile
          \State $ \mathcal{ T } \leftarrow \mathcal{T}_{i-1} $ \Comment{set outputs to the values corresponding to minimum residual.}
          \State $ \mathcal{ G }^* \leftarrow \mathcal{ G }^*_{i-1} $
          \State $ \mathsf{ N } \leftarrow \mathsf{ N }_{(i-1)} $
          \State \Return $ \mathcal{T} $, $ \mathcal{ G }^* $, $ \mathsf{ N } $.
        \end{algorithmic}
      \end{minipage}
    \end{tabular}
  \end{ruledtabular}
\end{table}
\endgroup

\begingroup 
\squeezetable
\begin{table}
  \caption{\label{algAutotuning}Orientation-recovery using a self-tuning kernel}
  \begin{ruledtabular}
    \begin{tabular}{c}
      \begin{minipage}{\linewidth}
        \begin{algorithmic}[1]
          \Statex \textbf{Inputs:}
          \Statexi Snapshots $ \mathcal{ M } = \{ \underline{ a }_1, \ldots, \underline{ a }_s \} $
          \Statexi Number of retained nearest neighbors $ d $ 
          \Statexi Number of datapoints in the least-squares fit, $ r $
          \Statexi Number of nearest neighbors for autotuning, $ n $
          \Statex
          \Statex \textbf{Outputs:}
          \Statexi Estimated quaternions $ \mathcal{ T } = \{ \tau_1, \ldots, \tau_{s} \} $,
          \Statexi Nearest-neighbor index matrix $ \mathsf{ N } $  
          \Statexi Least-squares residual $\mathcal{ G }^* $
          \Statex
          \State Compute the $ s \times d $ matrices $ \mathsf{ N } $ and $ \mathsf{ S } $ such that
          \begin{displaymath}
            \begin{aligned}
            N_{ij} &= \text{index of $j$-th nearest neighbor to snapshot $ \underline{ a }_i $},\\
            S_{ij} &= \lVert \underline{ a }_i - \underline{ a }_{N_{ij}} \rVert.
            \end{aligned}
          \end{displaymath}
          \State \Return $ \mathsf{ N } $
          \State Rescale the distance data by the $ n $-th nearest neighbors:
          \begin{displaymath}
            S_{ij} \leftarrow  \left( S_{i,N_{i,n }} S_{j,N_{j,n}} \right)^{1/2}.
          \end{displaymath}
          \State Compute the sparse transition probability matrix $ \mathsf{ P } $ using the algorithm in Table~\ref{algP} with inputs $ \mathsf{ S } $, $ \mathsf{ N } $, $ \epsilon $, and $ \alpha = 1 $.
          \State Solve the sparse eigenvalue problem $ \mathsf{ P } \underline{ \psi }_k = \lambda_k \underline{ \psi }_k $ for $ 0 \leq k \leq 9 $ and $ 1 = \lambda_0 < \lambda_1 \leq \cdots \leq \lambda_9 $.
          \State Solve the nonlinear least-squares problem 
\begin{displaymath}
    \mathcal{ G }( \{ c_{ijk} \} ) = \sum_{l=1}^{ r } \lVert \tilde{ \mathsf{ R } }_l^\mathrm{T} \tilde{ \mathsf{ R } }_l - \mathsf{ I } \rVert^2 + | \det( \tilde{ \mathsf{ R } }_l ) - 1 |^2 \text{    with    } [ \tilde{ \mathsf{ R } }_l ]_{ij} = \sum_{k=1}^9 c_{ijk} \psi_{lk}.
\end{displaymath}
          \State \Return $ \mathcal{ G }^* $
          \For{$ i = 1, \ldots, s $}
          \State Compute an approximate SO(3) matrix $ \tilde{ \mathsf{ R } }_i $ for snapshot $ \underline{ a }_i $
          \State Project $ \tilde{ \mathsf{ R } }_i $ to an orthogonal matrix
          \State Convert $ \mathsf{ R }_i $ to a unit quaternion $ \tau_i $
          \State \Return $ \tau_i $
          \EndFor
        \end{algorithmic}
      \end{minipage}
    \end{tabular}
  \end{ruledtabular}
\end{table}
\endgroup

\begingroup 
\squeezetable
\begin{table}
  \caption{\label{algP}Calculation of the sparse transition probability matrix $ \mathsf{ P }_\epsilon $ in Diffusion Map, reproduced from Paper~I for convenience.}
  \begin{ruledtabular}
    \begin{tabular}{c}
      \begin{minipage}{\linewidth}
        \begin{algorithmic}[1]
          \Statex \textbf{Inputs:}
          \Statexi $ s \times d $ distance matrix $ \mathsf{ S } $
          \Statexi $ s \times d $ nearest-neighbor index matrix $ \mathsf{ N } $
          \Statexi Gaussian width $ \epsilon $
          \Statexi Normalization parameter $ \alpha $
          \Statex
          \Statex \textbf{Outputs:}
          \Statexi $ s \times s $ sparse transition probability matrix $ \mathsf{ P } $
          \Statex
          \State Construct an $ s \times s $ sparse symmetric weight matrix $ \mathsf{ W } $, such that
          \begin{displaymath}
            W_{ij} = \begin{cases}
              1, & \text{if $ i = j $}, \\
              \exp( - S_{ik}^2 / \epsilon ), &  \text{if $ j = N_{ik} $}, \\
              W_{ji}, & \text{if $ W_{ij} \neq 0$}, \\
              0, & \text{otherwise}.
            \end{cases}
          \end{displaymath}
          \State Evaluate the $ s \times s $ diagonal matrix $ \mathsf{ Q } $ with nonzero elements $ Q_{ii} = \sum_{j=1}^s W_{ij} $.
          \State Form the anisotropic kernel matrix $ \mathsf{ K } = \mathsf{ Q }^{-\alpha} \mathsf{ W } \mathsf{ Q }^{-\alpha} $.
          \State Evaluate the $ s \times s $ diagonal matrix $ \mathsf{ D } $ with nonzero elements $ D_{ii} = \sum_{j=1}^s K_{ij} $.
          \State \Return $ \mathsf{ P }_\epsilon = \mathsf{ D }^{-1} \mathsf{ K } $ 
        \end{algorithmic}
      \end{minipage}
    \end{tabular}
  \end{ruledtabular}
\end{table}  
\endgroup

Next, we enter an iteration loop, where in the $ i $-th step the dataset
\begin{equation}
  \label{eqAIterate}
  \tilde{ \mathcal{ M } }_i = \VSTL( \mathcal{ M }_I; \sigma, \mathsf{ N }_{i-1} )
\end{equation}
is computed, and the algorithm in Table~\ref{algAutotuning} executed using $ \tilde{ \mathcal{ M } }_{i} $ as input data. The resulting quaternion estimates, nearest-neighbor indices, and least-squares residual are respectively designated $ \mathcal{ T }_i $, $ \mathsf{ N }_{i} $, and $ \mathcal{ G }_{i}^* $. 
Note that the residual $ \mathcal{ G }_{i}^* $ is a measure of the difference between the eigenfunctions obtained by embedding and the natural eigenfunctions (Wigner $ D $-functions) expected on the basis of symmetry (see Paper~I).

If, after an iteration $ i $, $ \mathcal{ G }^*_{i} $ is larger than the residual $ \mathcal{ G }^*_{i-1} $ encountered in the previous step, the  loop is terminated. Otherwise, the iteration is repeated using  $ \mathsf{ N }_{i} $ as an updated estimate of the true nearest-neighbor indices. Our final orientation (quaternion) assignment is the one corresponding to the minimum least-squares residual, reached in the iteration prior to the termination step. 

The empirical evidence in Sec.~\ref{secSimulatedData} clearly shows that, given a sufficiently large number of sample points, the scheme, applied only a handful of times and terminated using the value of $ \mathcal{ G }_i^* $ as a criterion, provides noise reduction sufficient for accurate orientation recovery at $ \text{MPC} = \Ord( 10^{-2} ) $.

A potentially fruitful way of interpreting mathematically the success of the process (which lies outside the scope of the present paper) would be to explore its connections with mutually reinforcing models (MRMs) for graph filtering \cite{ChenSafro10}. This type of model involves iteratively replacing vertices of graphs with weighted averages, whereby the vertex itself and its local neighborhood exert an influence on the vertex in the course of iterative updates. In certain applications, the iterative process in an MRM is terminated after only a small number of iterations. Both of these two features are present in the scheme presented here.

\section{Computational resources}
\label{appendixComputationalResources}
The calculations reported in this work were primarily performed on a Rocks cluster with 30~nodes, each 
consisting of two 2.5 GHz Quad-Core Intel Xeon CPUs with 16 GB RAM. Algorithms were usually 
implemented in \textsc{matlab r2009b} with the Parallel Computing Toolbox together with the \textsc{matlab} Distributed 
Server using up to 120 workers (parallel processes). For less intensive calculations, a Linux workstation 
with a 2.66 GHz Quad-Core Intel Xeon CPU, 32 GB RAM and/or a Mac~Pro $ 2 \times 2.8 $ GHz Quad-Core Intel 
Xeon CPUs, 10 GB RAM were used.  In Diffusion Map, by far the most CPU-intensive calculations are: (1) 
the determination of the Euclidean distances of snapshots; and (2) setting up of the sparse distance 
matrix of nearest neighbors. These calculations were performed in parallel using 100~workers. Such a 
distance calculation involving $ 2 \times 10^6 $ snapshots typically takes 7~hours for chignolin and 48~hours for adenylate kinase (ADK) in Paper~I. 
Other calculations, including the eigenvector determination and the estimation of the orientation matrices were performed on 
the Linux workstation in about 8~hours altogether. In total, the orientation determination for  $ 2 \times 10^6 $ 
snapshots requires 56~hours for noise-free ADK and 33~hours for noisy chignolin with 5~local-averaging 
iterations. Compiling a 3D diffraction volume consisting of a uniform Cartesian grid was implemented in 
parallel code, with an execution time of less than three hours for two million ADK
diffraction snapshots using 80 workers. Diffraction patterns and cryo-EM images were simulated on the 
cluster and on the Mac Pro. The GTM and phasing algorithms were performed 
on the Linux workstation and/or the Mac~Pro. The \textsc{chimera} package~\cite{Chimera} was used to visualize electron 
density maps. 

%

\end{document}
%